# Benefits of Shifting Passenger Traffic from Air to Rail

## A Case Study of California High-Speed Rail

Kaijing Ding, Lu Dai, Mark Hansen
Institute of Transportation Studies
University of California, Berkeley
Berkeley, CA, USA
kaijing@berkeley.edu, dailu@berkeley.edu, mhansen@ce.berkeley.edu

*Abstract*— **This study provides a method to quantify the benefit of shifting passenger traffic from air to high-speed rail (HSR) from the perspective of flight delay cost reduction. We first estimate the number of flight reductions for airport origin and destination pairs based on the HSR ridership forecast provided in the California High-Speed Rail 2020 Business Plan, and then distribute these flight reductions to quarter hours. After that, Lasso models are applied to estimate the impact of the reduced queuing delay of SFO, LAX and SAN on the arrival delay of national Core 29 airports. Finally, these delay reductions are monetized using aircraft operating cost per hour and the value of passenger time per hour. We apply several different variations of this approach, for example, considering delay at all 29 Core airports or just the major California airports, different scenarios for future airport capacity and flight schedules, and different percentiles of a probabilistic forecast for future HSR ridership. We ultimately arrive at delay cost savings of $51-88 million 2018 dollars in 2029 and 235-392 million 2018 dollars in 2033.**

## I. INTRODUCTION

California has two of the ten busiest airports in the country by both flights and passengers [1]. The route from Los Angeles International Airport (LAX) to San Francisco International Airport (SFO) is found to be the 9th busiest domestic route in the world [2]. Data from the Bureau of Transportation Statistics (BTS) [1] further show that the inter-regional travel between the five primary airports in the Los Angeles Basin and the three primary airports in the San Francisco Bay Area, which carries over 13 million people per year, is the highest in the U.S. The California airports are also becoming increasingly capacity-constrained [3,4] and are experiencing worse delays. SFO, which has 23% of flights delayed by at least 15 minutes, ranks the third worst delayed airport in the country [1].

Though there was a substantial downturn in travel demand in 2020 due to the pandemic, travel demand and operations levels are rebounding to their prior growth trajectory and bringing back congestion mitigation challenges. While the airports have been actively seeking solutions to increase capacity (e.g., expanding terminals, adding new gates [5]), another option is that passengers may eventually shift to the California High-Speed Rail (CAHSR), which is currently under construction. The completion of the Initial Phase is projected to provide service in the 171-mile segment from Merced to Bakersfield in the Central Valley starting from 2029 [6]. Under current plans, Full Phase 1 will connect Anaheim and Los Angeles with San Francisco via the Central Valley in 2033, covering a total distance of 380 miles [6]. According to the 2015 Interregional Transportation Strategic Plan [7], the highest priority of transportation development for the corridor between the San Francisco Bay Area and Los Angeles has been given to the CAHSR. The HSR is viewed as a transformative mobility option to reduce traffic demands on California roads and airports, which is considered to be a major quantifiable benefit. The need to better understand and quantify the benefits of shifting traffic demands from other modes to HSR has been of immediate concern to the public and has attracted increasing research attention in recent years.

A widely used method to quantify the benefits of HSR in practice is the Cost-Benefit Analysis, which evaluates the potential social and economic impacts of the proposed project [8]. The costs considered usually include infrastructure costs, operating costs and external costs (e.g., land resumption, barrier effects, noise), while the benefits usually include ticket revenues, travel time savings, pollution reduction, reliability improvement, safety improvement, and regional economic development [9-12]. Many methods have been applied to assess these costs and benefits, including but not limited to game engineering methodology [13], multinomial and mixed logit models [14], optimization problem formalization [15].

Among the variety of benefits, this study focuses on the benefit received by air transportation from HSR operation. A few papers have studied the complementary effects between HSR and air. Albalate et al. [16] provided evidence that HSR can provide feeding services to long-haul air services in hub airports, based on an empirical analysis of the European market. López-Pita and Robusté [17] presented the possible effects of High-Speed Rail in reducing slots needed at Madrid Airport by comparing them to other studies carried out in Paris-Charles de Gaulle airport and Frankfurt airport. Zhang et al. [18] found that air-HSR integration has significantly positive impacts on airport enplanement at primary hub airports in East Asian regions. Li and Rong [19] assessed the impact of the HSR network on the resilience of the air transport network against random failure and malicious attack.

For the case study of California HSR, the current method for monetizing highway and airport traffic reduction benefits is the ***Equivalent Capacity Analysis (ECA)*** used in the 2020 Business Plan [5]. This analysis compares the cost of adding capacity to the existing state highway network and airports to accommodate the passenger traffic that is forecast to be diverted to high-speed rail. For air travel, the analysis first estimates the total HSR people-carrying capacity based on the potential train frequency, train seating capacity, and the average load factor. With the assumption that 20% of high-speed rail capacity would be served by air travel if HSR were not built, it then distributes the required air capacity between the airports based on the market share of airports according to the overall flight patterns in California. Next, the needed number of gates, runways and parking lots at each airport is derived based on the aircraft seating capacity, load factor, gate capacity, and runway capacity. Then the airport expansion cost is calculated, including the construction cost of runways, gates, parking lots and other expenses. The results of this analysis indicate that a comparable airport infrastructure investment would require the construction of additional 91 gates and 2 runways across the state, which leads to a total cost of 23.9 billion 2018 dollars. The final result is a lump sum of the estimated airport expansion cost scheduled for each year until the 2033 completion date.

Despite its advantage of being simple and easy to understand, the ECA method did not consider some important factors. First, the ECA uses the planned HSR capacity, instead of the forecasted HSR demand, to estimate the benefit. The benefits are highly likely overestimated because the planned capacity may include excess capacity, and thus the true demand is expected to be less than the capacity. Second, the current method only considers the benefits of the air passengers shifting to HSR from the perspective of comparable highway and airport capacity upgrade cost savings. It fails to capture a broader scope of benefits of HSR, for example, the travel time of HSR is known to be more punctual and reliable than air [20]; HSR has a lower externality cost than air travel and automobiles in terms of accidents and fatalities, noise, congestion and time, as well as air quality [11]; HSR can also improve the accessibility of the region in general [21]. Third, the potential benefits to the aviation industry, and the National Airspace System (NAS) on a broader scale, were not captured. HSR will attract part of the current air passengers and may lead to a reduction in flight operations and relieve the current congested airspace and airport runways, bringing delay cost savings to the airline and air passengers.

This study provides an alternative and complementary method to quantify the air traffic reduction benefits induced by HSR operations, by focusing on flight delay reduction. The contributions of this study are: (1) This is the first work to apply queuing theory and machine learning models to quantify the potential impacts of HSR on air transportation, using a wide range of datasets. (2) We use the HSR ridership (demand) forecast instead of planned HSR capacity to make the benefit estimate, which is less likely to overestimate the benefits and can reflect the future scenario better. (3) Our study is unique in its capability to convert the impact of air passenger diversion to flight delay reductions, and further monetize them to estimate the benefits of shifting passengers from air to HSR. (4) Our approach is able to capture a broader scale of benefits both within and beyond California by modeling the flight delays for individual airports and the NAS, while controlling for other factors including terminal conditions, en route weather, wind, traffic volume, and special events [22]. (5) This analysis is entirely based on open-source data and is easily adaptable when more detailed and accurate ridership forecast, airport capacity, and flight schedule information become available. (6) We develop methods for forecasting the quarter-hour level air demand and capacity, with and without HSR impact, which can be adapted to support other research endeavors that require future demand and capacity forecast at a fine level.

The rest of the paper is organized as follows. We introduce the various data sources in Section II and provide details about how we fuse them together to forecast future air capacity and demand in Section III. Section IV introduces the methodology for evaluating the impact of passenger shifting to HSR on the airport delay of Californian airports as well as national airports. Finally, we discuss the pros and cons of our proposed approach and compare its results with the one obtained from the ECA method in Section V.

## II. DATA SOURCES

In this study, we combine a wide range of data sources, mainly including two groups. The first group is the flight operation datasets used for air capacity/demand forecast and delay estimate.

The Aviation System Performance Metrics (ASPM) dataset by Federal Aviation Administration (FAA) provides quarter-hour level historical airport scheduled arrivals and departures, as well as capacity information determined by visibility, winds, and fleet mix. We collected more than 20 million quarter-hour observations over the analyzed airports (including the Core 30 airports [23] except for Honolulu International Airport) from 2010 to 2019. We also collect the ASPM flight-level dataset in order to calculate the traffic share between the airport OD pairs in California.

- The FAA Terminal Area Forecast (TAF) contains both historical records and future forecasts of the annual number of flight operations for the analyzed airports that we will focus on.
- The FAA Airport Capacity Profile gives the model-estimated future hourly rate of arrivals and departures under certain runway configurations, flight rules, and separation standards. This dataset is available for the Core 29 airports only.
- The on-time performance dataset extracted from the Bureau of Transportation Statistics (BTS) TranStats data library, contains the positive arrival delay against schedule (in minutes), canceled flight indicator, diverted flight indicator, and the diverted arrival delay against schedule (in minutes) of each aircraft. Our flight delay metric – daily average arrival delay (in minutes per flight) is computed based on Dai et al.'s work [22] using this dataset. Only positive delay against schedule (not greater than 10 hours) is considered, where (negative) delays of early arrivals are counted as zero. Flight cancellation is counted as a 120-minute delay and the diverted arrival delay against schedule is used for diverted flights. Flights are assigned to days based on their scheduled arrival time.

The second group of datasets includes HSR ridership forecasts, historical mode share information, and air carriers data, which are used to forecast the flight reductions resulting from passengers shifting to HSR.

- The CAHSR 2020 Business Plan Ridership and Revenue Forecasting Technical Report [24] gives annual HSR ridership forecasts for each major market for the year 2029 (the completion year of the Central Valley segment) and 2033 (the completion year of the full Phase 1), as shown in the third and the fourth columns of Table I. These point estimates, which will be referred as "base run" results in the following sections, are ridership estimates using the base input values (HSR train frequency, ticket price, etc.). The CAHSR Ridership and Revenue Model Documentation [25] also give range estimates, i.e., different quantiles (minimum, 1%, 10%, 25%, median, 75%, 90%, 99%, maximum) of ridership based on Monte Carlo simulation.
- The CAHSR Ridership and Revenue Model Documentation [14] includes the result from the 2012-2013 California Household Travel Survey (CHTS), which shows the before-HSR-time mode share of auto, air, and conventional rail (CVR) for the major markets. These numbers are shown in the 5th-7th columns of Table I.
- The Air Carriers: T-100 Domestic Market (All Carriers) data from the BTS website is used to learn the airport share of each market. The BTS Air Carriers: T-100 Domestic Segment (All Carriers) data is used to find the average passenger load for each airport OD pair and transform the number of diverted passengers into the number of flights eliminated. For both datasets, the 2019 full year data is applied since it excludes the abnormal pandemic effects and should be more reliable.

## III. Air Travel Forecast

Our proposed method estimates the benefits of shifting passenger traffic from air to rail, from the perspective of cost savings from flight delay reduction. Toward this end, we need to forecast what the air capacity and demand will be once the CAHSR is in operation, and to what extent the HSR may affect the flight operations. Moreover, we seek to derive these forecasts on a fine granularity – quarter-hourly level – in order to leverage the sophisticated delay model developed by Dai et al. [22].

### A. Air Capacity Forecast

Two kinds of air capacity estimates are used in this study. The first one is a conservative estimate, which assumes the capacity to remain the same as in 2019 for future years. These capacity estimates are directly obtained from the ASPM dataset, and will serve as a baseline to represent the worst-case scenario, where there is no future airport capacity improvement.

We also consider a future capacity scenario based on the Airport Capacity Profiles published by FAA. Specifically, we first match the quarter-hour data record from the ASPM dataset with the hourly arrival/departure rate provided in the Capacity Profiles, according to the airport, runway configuration, and weather (ceiling, visibility) information. Then the quarter-hour airport arrival/departure rate (AAR/ADR) is forecasted to be the future hourly rate divided into four (quarter-hours). Note that the Capacity Profiles does not specify the year when the improvements will be implemented, but it is reasonable to assume that they will be in place by 2029. Also, it is likely that, should demand grow as forecasted, additional improvements would be made after 2029. In this analysis, we assume that the 2029 and 2033 capacities are the same and comply with the capacity improvements estimated in the Capacity Profiles.

### B. Air Demand Forecast

In this section, we aim to forecast future flight scheduled demand on a quarter-hour level by integrating the ASPM dataset and the TAF data. However, there are two difficulties in utilizing the TAF demand forecast. First, the TAF dataset only gives the annual number of operations and does not provide data at a quarter-hour level as the ASPM flight schedule. The second problem lies in the discrepancy between the ASPM and TAF datasets due to the metric difference. The TAF data counts the itineraries, while the ASPM data considers OAG scheduled operations.

Our strategy to tackle the above issues is first to find the relationship between the annual numbers of the TAF data and ASPM data, then scale up the ASPM quarter-hour scheduled demand with the annual number ratio to obtain future estimates. For each airport, the historical annual number of OAG scheduled arrivals, $Q_{ASPM}^{arr}$ , and departures, $Q_{ASPM}^{dep}$ , is calculated by aggregating the ASPM quarter-hourly records over the year. Then a linear regression was applied to find the relationship between the annual number of operations from TAF, $Q_{TAF}$, and the annual OAG scheduled operations from ASPM ($Q_{ASPM}^{arr}, Q_{ASPM}^{dep}$). With all the 77 airports tracked by the ASPM over the ten years from 2010 to 2019, the data yields 770 observations, and the estimation results are as follows:

$$Q_{ASPM}^{arr} = 0.505 \cdot Q_{TAF} - 27649 \quad (1)$$

$$Q_{ASPM}^{dep} = 0.506 \cdot Q_{TAF} - 27712 \quad (2)$$

The $R^2$ values for both regression models are very high – 0.952. With these regression models, we can convert the TAF forecast into equivalent annual values of ASPM arrivals and departures for future years. Taking 2019 as the base year, its quarter-hour scheduled arrivals and departures can be scaled up with the ratio of ASPM annual operations in the future year to that in 2019. This procedure is repeated for both arrivals and departures, all airports of concern, and all quarter hours of 2029 and 2033. We keep the non-integer results, since the queuing model used later is based on a continuum approximation. Also, as discussed below, we consider scenarios in which the future demand is "de-peaked" from that obtained using the method above.

### C. Shifting Traffic from Air to HSR

In this section, we consider the HSR impact on air transportation by estimating flight reductions for each airport OD pair based on the HSR ridership forecast provided in the 2020 Business Plan.

#### 1) Annual Diverted Passengers by Market

The main mode choice model used in 2020 Business Plan ridership forecast rejected the nested structure where air, HSR and conventional rail are nested under common carrier and then compete with car; and decided on the alternative structure with car, air and rail as three parallel choices and HSR and conventional rail nested in rail. In other word, there is no evidence that air passengers for the market of interest might be more or less willing to take HSR than car drivers or passengers. Therefore, this study assumes that the number of diverted passengers from each mode to HSR is proportional to their current mode share, the number of diverted passengers from air to HSR, $P_m^{yr}$, for market $m$, for the future years ($yr = 2029, 2033$) is calculated:

$$P_m^{yr} = N_m^{yr} \cdot \alpha_m \quad (3)$$

where $N_m^{yr}$ is the total HSR ridership of the market $m$ at the given year $yr$ obtained from the 2020 Business Plan, and $\alpha_m$ is the pre-HSR airline mode share of market $m$ obtained from 2012-2013 CHTS. The result for the MTC market is shown in Table I.

The total number of diverted passengers from air to HSR for all markets is 906K and 2,557K for 2029 and 203, accounting for 5.6% and 7.2% of all forecasted HSR ridership in the corresponding year. The HSR ridership range estimates provided in [25] are also included in this study to reflect the uncertainty in our benefit estimates. Unless otherwise specified, the following steps are all performed for the different quantiles of HSR forecasts; while the tables, if not specifying the quantiles, only show the numbers using base run estimates due to limited space.

*2) Annual Diverted Passengers by Airport Pair*

Next, we decompose the market-wise passenger diversions for each airport OD pair. For this purpose, we utilized the BTS T-100 Domestic Market data and grouped the passenger volume by origin and destination airport pairs for each market $m$, then calculated the traffic shares of each OD pair, $\beta_{od}^m$, among airports. We assume that annual diverted passengers from air to HSR follow the same traffic distribution probability. Thus, the number of diverted passengers $P_{od}^{yr}$ for each airport pair $od$ in a future year $yr$ is estimated as:

$$P_{od}^{yr} = P_m^{yr} \cdot \beta_{od}^m \quad (4)$$

*3) Flights Reduction by Airport Pair*

Next, we transform the diverted passenger estimates to the number of eliminated flights for each airport pair. We used the BTS T-100 Domestic Segment data to compute the average number of passengers per flight, $\gamma_{od}$, which is calculated as the ratio of the total number of passengers to the total number of flights for the origin and destination airport pairs of concern. We then estimate the flight reduction $F_{od}^{yr}$ by airport pair $od$ in the future year $yr$ to be:

$$F_{od}^{yr} = \frac{P_{od}^{yr}}{\gamma_{od}} \quad (5)$$

TABLE I. ANNUAL DIVERTED PASSENGERS FROM AIR TO HSR BY MARKET

| Market | | Annual HSR Ridership Forecast (Millions) | | Before-HSR-Mode Time Share | | | Number of Diverted Passengers from Air to HSR (Thousands) | |
|---|---|---|---|---|---|---|---|---|
| | | *2029* | *2033* | *Auto* | *Air* | *CVR* | *2029* | *2033* |
| MTC | MTC | 1.8 | 2.2 | 99% | 0% | 1% | 0 | 0 |
| MTC | SCAG | 2.4 | 6.4 | 69% | 30% | 1% | 720 | 1920 |
| MTC | SJV | 4.3 | 4.6 | 99% | 0% | 1% | 0 | 0 |
| MTC | Others | 2.3 | 2.7 | 99% | 0% | 1% | 0 | 0 |

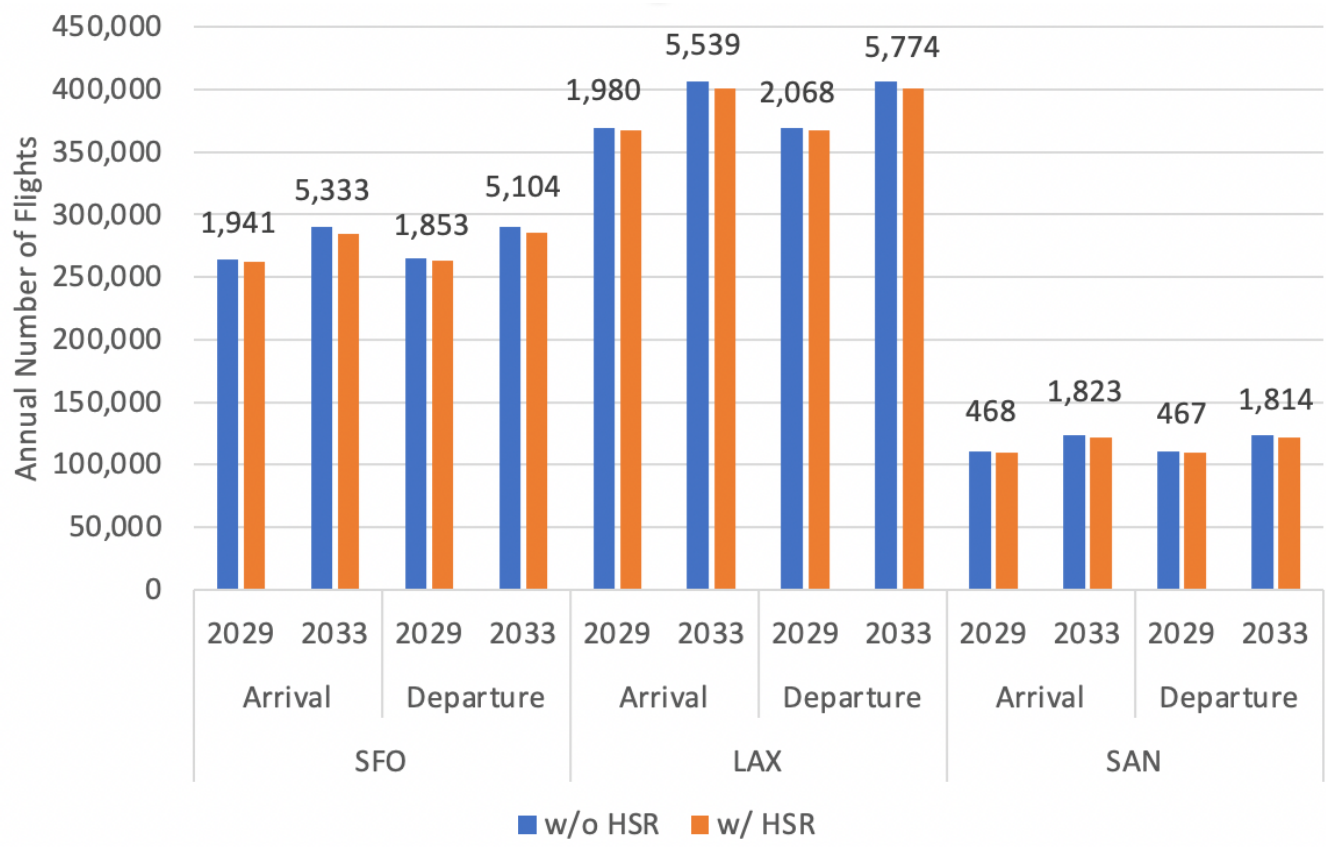

Figure 1. Airport Annual Number of Arrivals and Departures Forecast with and Without HSR Impact

*E. Air Demand Forecast, With HSR Impact*

Given annual flight reductions between airport pairs resulting for HSR, we now translate these annual reductions into changes in quarter-hour air traffic arrivals and departures from the baseline air demand forecasts obtained in Section B. We assume that the quarter-hours with more arrivals (or departures) from a certain origin (or to a certain destination) in 2019 are more likely to experience flight reduction for that origin (or destination) in future years. Accordingly, we randomly choose the quarter-hours that will have flight reductions for a given directional airport-pair based on the temporal distribution of flights for that pair in 2019. We then sum up the number of flight reductions from all origin (or destination) airports to the airport of concern to obtain the quarter-hour arrivals (or departures) of this airport.

Our demand forecast does not consider the capacity constraints and may result in extremely large quarter-hour demand rates. We followed the current practice in the airline industry to de-peak the demand within the hour to increase schedule reliability [26]. Specifically, the new quarter-hour demand rate is calculated as the mean of the four quarter-hour demand rates of that hour. Fig. 1 illustrates the aggregated forecast result of the annual number of arrivals and departures by the future year. The annotated numbers on the top show the difference between the air demand forecast with and without HSR impact.

## IV. FLIGHT DELAY PREDICTION

*A. Methodology*

The analysis above has yielded forecasts of future airport capacity, and flight demand schedules with and without diverting passenger traffic from air to HSR. In this section, we aim to translate the HSR impacts on air travel into flight delay cost savings at California airports, and the NAS. Particularly, we construct four scenarios from the air travel forecasts results: the ***Baseline*** scenario with baseline air traffic demand (without de-peaking) but no capacity improvement (remain the 2019 capacity level), the ***Cap***acity only scenario with baseline air traffic demand and capacity improvement, the ***De-peak***ing scenario with de-peaking air traffic demand but no capacity improvement, and the capacity and de-peaking (***Cap & De-peak***) scenario with both capacity improvement and de-peaking air traffic demand. The following analysis will be performed on all four scenarios. We will first employ a simple queuing model to estimate the arrival and departure queuing delays at each airport based on the air capacity and demand forecasts. Then we adapted the predictive delay model proposed by Dai et al. [22] to study the impact of queuing delays on flight operation performance at SFO, LAX, and SAN, as well as the NAS. Furthermore, we use the results to evaluate the impact of flight reductions resulting from HSR traffic diversion on the flight delay.

### *B. Queueing Delay*

A basic building block of the delay models used in this research is the deterministic queuing diagram. In this study, we employ the cumulative input-output (I/O) queueing diagram to compute the queueing delays based on the forecasted quarter-hour flight schedules and airport capacity values for both arrivals and departures.

Figure 2 illustrates a queuing diagram of flight arrivals at an airport on a certain day, with the orange curve depicting cumulative capacity constrained arrivals and the blue curve showing cumulative scheduled arrivals. The area between the two curves is the total deterministic queuing delay (in quarter-hours) for all flights arriving at this given airport on the given day, which can be segmented for each quarter-hour to calculate flight average queuing delay. We further aggregated and averaged the quarter-hour queuing delay for four time periods of the day: 0:00-6:00, 6:00-12:00, 12:00-18:00, and 18:00-24:00 so that this feature is in accord with Dai et al. [22]'s work and can be used for the next step.

### *C. Predictive Modeling*

In this section, we leveraged the comprehensive predictive models proposed by Dai et al. [22] to model the flight delay in the three California major airports (SFO, LAX, and SAN) and the NAS. Dai et al. [22] applied various machine learning algorithms to model the NAS delay for the period from 2010 to 2019 using a broad scope of spatial-temporal features, including queuing delays, terminal conditions, en route weather, wind, traffic volume, and special events. We adapted this feature matrix for our study by updating the queueing delay features, scheduled arrival features, and including the new unserved demand features. We also further expanded the response variables to be airport specific. The response variable in their model is the daily average arrival delay (in minutes per flight), which is computed based on average positive delay against schedule for all scheduled arrivals into the Core 29 U.S. airports, adjusted to include flight cancellations. Besides this nationwide delay metric, we also include the daily average flight arrival delay of SFO, LAX, and SAN for our study. We run the models independently for these four response variables to capture the effects at different scales. Table II gives a description and the summary statistics of these four different response variables. SFO appears to have the highest delay and variability, while LAX and SAN have a more moderate level of delay close to, about the same level as the national average.

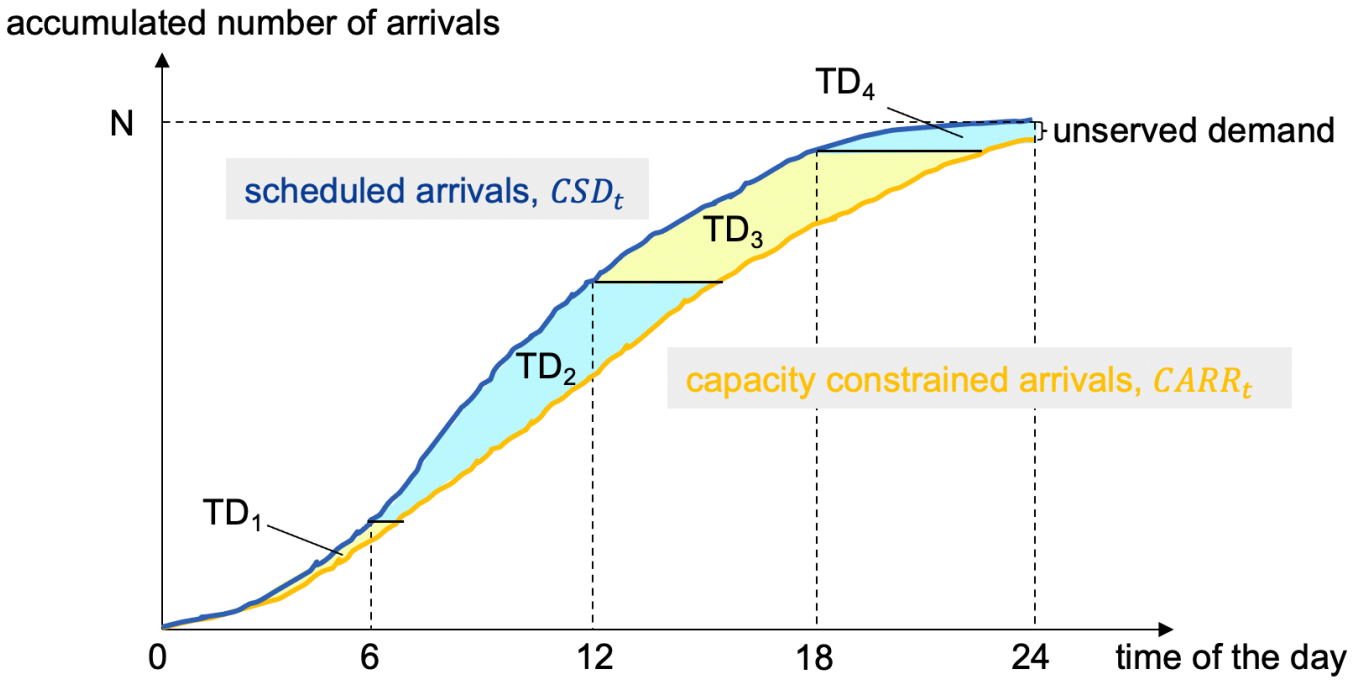

Figure 2. Queuing diagram

TABLE II. DESCRIPTION OF THE RESPONSE VARIABLES

| | **Mean (min)** | **Standard deviation (min)** | **Minimum (min)** | **Maximum (min)** |
|---|---|---|---|---|
| NAS (Core 29 U.S. airports) | 14.08 | 7.19 | 2.62 | 55.51 |
| SFO | 20.76 | 18.86 | 1.22 | 132.90 |
| LAX | 13.28 | 7.22 | 2.33 | 79.42 |
| SAN | 11.44 | 6.41 | 0.59 | 67.34 |

Among the seven machine learning models they experimented, Dai et al. chose Lasso as the final model considering interpretability, computational efficiency, and model performance. Therefore, we opted for the Lasso model as well. We removed the same outlier observations – 42 days that have detected outlier features and 20 days with daylight savings time transition – as in Dai et al.'s work. The mean absolute errors (MAE) of these models with the optimal fine-tuned hyperparameters on the exact same testing set are 2.67, 4.82, 3.23, and 3.02 minutes/flight for the NAS, SFO, LAX, and SAN model, respectively. These results show how accurately the models, when trained on a historical data set of days between 2010 and 2019, predict delays on a testing set of other days, which were not used in training, of days in the same time period. The results show that the performance of the Lasso model varies for different response variables. The MAE is higher for SFO than LAX and SAN. The model performance is similar for the training and testing set, which means these models are not overfitting. We thus can generalize them to a new dataset where the queuing delay features are manipulated to reflect the HSR impact.

### *D. Scenario Analysis*

We now use the above models to predict how average arrival delay, for either the major California airports or the NAS, will change after the introduction of HSR, keeping all other factors (e.g., capacities) equal. Specifically, we first updated *four* sets of features in the feature matrix to reflect the future year condition impacted by HSR: airport-specific time-of-day average arrival queuing delay, airport-specific time-of-day average departure queuing delay, airport-specific time-of-day scheduled arrival counts, and airport-specific daily unserved demands. In total, values of 377 variables (3 feature sets * 29 airports * 4 time periods of the day + 1 feature set * 29 airports) will be updated to construct the *four* different scenarios we defined in Section IV.A. Next, we applied the trained model to the new feature matrix and obtained future delay predictions. Finally, for each response variable, each future year, each scenario, each HSR ridership quantile, we calculated the difference between delay prediction with and without HSR impact, $\delta_t$, using Equation (6) [27]. The difference represents the delay savings induced by the factor of change – HSR operations—on day $t$.

$$\delta_t = \hat{f}(\boldsymbol{X_t}) - \hat{f}(\boldsymbol{x_{tc}}, \boldsymbol{X_{t\backslash c}}) \tag{6}$$

where $\hat{f}(\boldsymbol{X_t})$ is the predicted average arrival delay on day $t$ provided by the selected model based on the feature matrix without HSR impact; $\boldsymbol{x_{tc}}$ are the amended flight queuing delays after taking HSR traffic diversions into account; and $\boldsymbol{X_{t\backslash c}}$ are the remaining unchanged features. Thus $\delta_t$ represents the difference in predicted delay at a California airport, or the NAS as a whole, for a day whose weather and wind features match those for the corresponding day in 2019. The queueing features of this future day are based on 2019 demand scaled to predicted traffic growth, which in some scenarios is de-peaked, and 2019 capacity, which in some scenarios is adjusted based on anticipated capacity enhancements.

We repeat this procedure for the four response variables (SFO, LAX, SAN, and the NAS), two future years (2029 and 2033), four scenarios defined in Section IV.A, and with or without HSR impact, under different HSR ridership estimates for various quantiles. Our focus is on delay savings resulting from HSR operation. Thus, for each airport or the NAS, each capacity/demand scenario, each HSR ridership estimate quantile, and two future years, the difference in predicted delay (i.e., delay savings) with and without HSR impact is calculated. Figure 3 compares the mean value of predicted daily average delay savings induced by HSR operation using the different HSR ridership quantiles and under different scenarios. The vertical axis represents the delay savings computed by subtracting the with-HSR delay predictions from the without-HSR delay predictions. Each subplot shows the results for the NAS, SFO, SAN, and LAX (clockwise), with lines indicating the baseline, capacity improvement, and de-peaking scenarios in 2029 and 2033. Note that although each plot has eight lines, only four are visible; this is because scenarios with and without de-peaking have almost identical delay savings. In addition, we plot the 25% and 75% percentile of the predicted daily average delay savings for the 2033 Cap & De-peak scenario to show the variability in delay savings. The insights drawn from the results are summarized.

1) The delay savings of all scenarios, including the cases investigated in Figure 3, are positive. This indicates that the introduction of HSR will improve the aviation system performance. HSR, as a new mode choice, will lead to air passengers diversion, reduce the number of flight operations, and thus ease the airspace congestion.

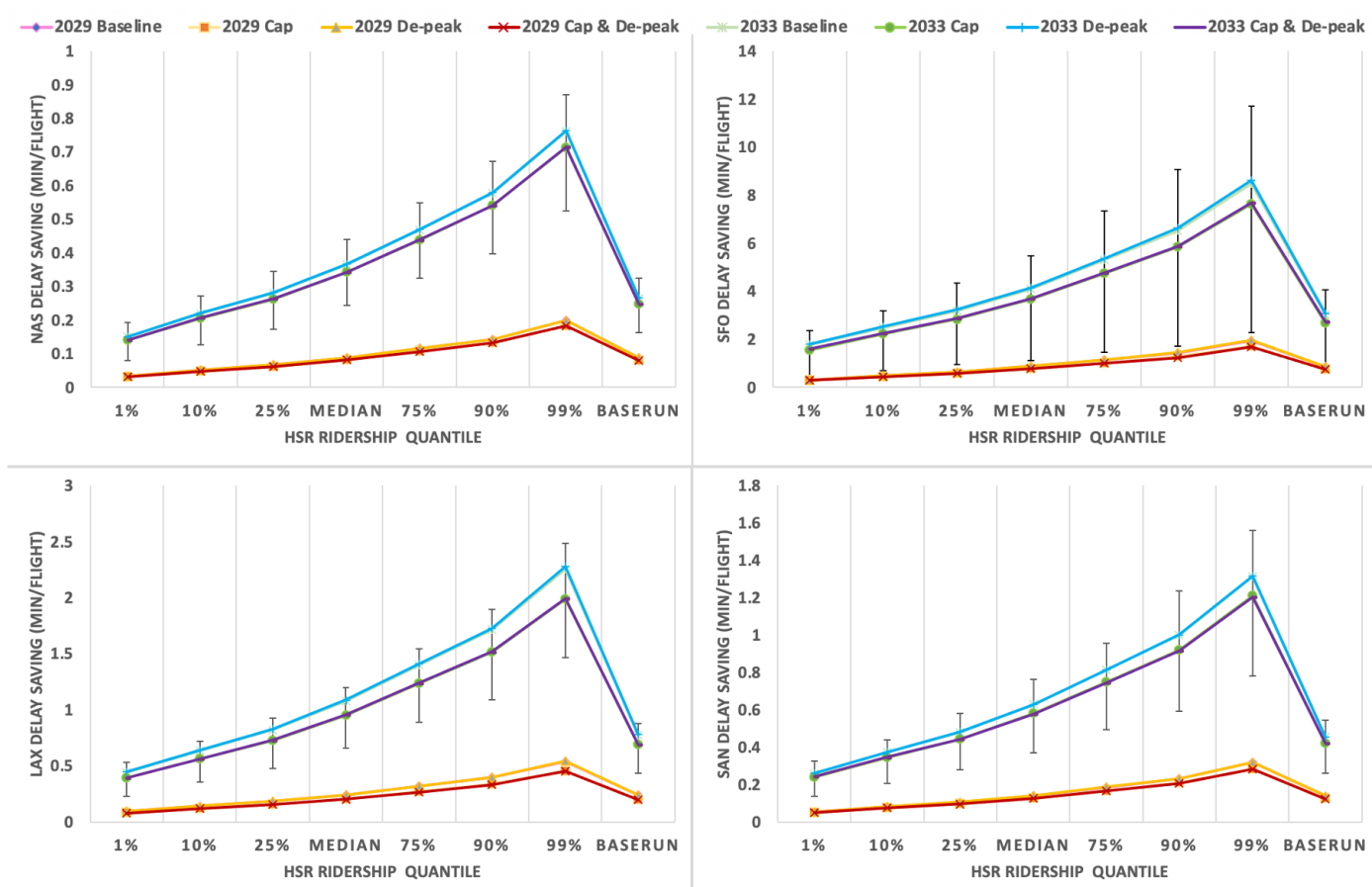

Figure 3. Comparison of Mean of Predicted Daily Average Delay Saving Given by Four models

2) SFO has both the highest delay saving and the largest variability of delay saving, LAX the second, SAN the third, and NAS the least. This is in the same order as the order of mean and variance of the original response variables (shown in Table II), which indicates that in general, the more congested airports may gain more delay savings from the introduction of HSR.

3) There is wide variability in delay savings from day to day. In the most extreme cases at SFO, delay savings from HSR exceed 28 minutes per flight. These reflect days with very low capacity, which when combined with unattenuated traffic growth, result in an extremely high baseline delays than can be substantially reduced by eliminating a relatively small number of flights. Also, delays on such days would manifest in large numbers of flight cancellations (recall that our delay metric includes such cancellations).

4) Airport capacity improvements would decrease flight delays, and also diminish the delay reduction benefit of HSR. However, demand de-peaking, though would decrease flight delays, does not have an obvious impact on the delay saving of HSR (lines of de-peak and non-de-peak scenarios mostly overlap in Figure 3).

5) Delay savings increase from 2029 to 2033, as the travel demand increases, California airports become more congested, and the HSR is able to divert more passengers.

## V. Delay Monetization

In the final step, we want to monetize the delay reduction so that the benefits of delay savings can be compared with infrastructure investment costs, as calculated in the ECA method. We perform our benefit cost savings for the NAS, LAX, SAN, and SFO models, which we identify with the subscript *model.* We define the total benefit $B_{model}^{yr}$ as the sum of the airline operation cost savings $AS_{model}^{yr}$ and monetarized passenger time savings $PS_{model}^{yr}$, and calculate them with the following formulas.

$$AS_{model}^{yr} = \delta_{model}^{yr} \times F_{model}^{yr} \times OC^{yr}/60$$

$$PS_{model}^{yr} = \delta_{model}^{yr} \times E_{model}^{yr} \times VOT^{yr}/60$$

$$B_{model}^{yr} = PS_{model}^{yr} + AS_{model}^{yr}$$

where $\delta_{model}^{yr}$ is the delay saving predicted in the last section, in minutes per flight; $F_{model}^{yr}$ and $E_{model}^{yr}$ are the total number of flights and the total number of enplanements respectively, given by the TAF dataset; $OC^{yr}$ and $VOT^{yr}$ are the aircraft operating cost per hour and the value of passenger time per hour, respectively, given by [28]. They are also adjusted by inflation rate according to national Consumer Price Index published by U.S. Bureau of Labor Statistics to be in the unit of 2018 dollars in order to be in the same units as the results of ECA methods.

Though the national average arrival delay is less impacted by the introduction of HSR than the three major Californian airports as shown in Fig. 3, the NAS has more aircraft affected. As a result, the sum of cost savings of the three individual airports has the same magnitude as the cost savings of the Core 29 airports (national savings). For the capacity improvement and de-peaking scenario in 2029, the delay cost saving for the national Core 29 airports is $92 million, and that for the three major Californian airports is $52 million. This rough consistency of the two estimates indicates that the spillover effects of the decrease in the arrival delays of California airports on the rest of the NAS are not so large as to dramatically change the delay cost saving results based on California airports only.

As shown in Table III, we also estimated and compared the delay cost savings under different ridership forecast scenarios,. The median scenario gives an almost identical estimate as the one given by the base run scenario. The cost savings under the 1%, 10%, 25%, 75%, 90%, 99% quantile of HSR ridership forecasts are about 40%, 60%, 75%, 130%, 160%, 200% as much as those under the median scenario. Thus, the difference in ridership between the 1% and 99% quantile is associated with roughly a five-fold difference in delay savings benefit.

The models developed for different delay measures (i.e., considering national Core 29 airports, or only the three major Californian airports), different scenarios (i.e., considering capacity improvement and demand de-peaking or not), and different ridership forecasts, lead to a range of cost-saving estimates. Although demand de-peaking can reduce delay (as shown in Section IV.B), it does not greatly impact the difference in delay cost before and after the operation of HSR. For example, the cost savings under the baseline and de-peak scenarios for 2029, using the base run HSR ridership forecasts are both $72 million. On the contrary, though the capacity improvement does not improve delay as much as demand de-peaking does, it has a more substantial impact on the delay cost savings. For example, the cost savings of baseline ($72 million) and capacity improvement ($66 million) scenario of 2029, for the NAS using base run HSR ridership forecasts differ by $6 million.

To summarize our benefit estimates, we focus on the NAS model since it includes delay savings from throughout the nation. We assume the capacity upgrades since these are likely to have occurred by 2029. We assume demand de-peaking, though as mentioned above this makes little difference in delay reduction. Finally, for the HSR ridership forecast, we consider the range between 25% and 75%. Based on these assumptions, we estimate delay cost savings of $51-88 million in 2029 and $235-392 million in 2033.

TABLE III. ESTIMATED DELAY COST SAVINGS UNDER DIFFERENT RIDERSHIP FORECAST QUANTILES (MILLION DOLLARS)

| | | 2029 | | | | 2033 | | | |
|---|---|---|---|---|---|---|---|---|---|
| | | Baseline | CAP | De-peak | CAP & De-peak | Baseline | CAP | De-peak | CAP & De-peak |
| National | 1% | 28 | 26 | 28 | 26 | 134 | 126 | 135 | 126 |
| | 10% | 42 | 39 | 42 | 39 | 196 | 184 | 197 | 185 |
| | 25% | 55 | 51 | 55 | 51 | 250 | 234 | 251 | 235 |
| | median | 73 | 68 | 73 | 67 | 327 | 306 | 328 | 306 |
| | 75% | 95 | 89 | 95 | 88 | 419 | 392 | 420 | 392 |
| | 90% | 118 | 109 | 118 | 108 | 516 | 483 | 518 | 483 |
| | 99% | 163 | 151 | 164 | 151 | 680 | 638 | 682 | 639 |
| | baserun | 72 | 66 | 72 | 66 | 236 | 221 | 237 | 222 |
| Sum of SFO, LAX, SAN | 1% | 16 | 15 | 16 | 15 | 96 | 85 | 97 | 87 |
| | 10% | 25 | 23 | 25 | 22 | 136 | 121 | 138 | 123 |
| | 25% | 33 | 29 | 33 | 29 | 174 | 155 | 177 | 157 |
| | median | 45 | 39 | 45 | 39 | 226 | 202 | 229 | 203 |
| | 75% | 58 | 51 | 59 | 50 | 293 | 261 | 295 | 262 |
| | 90% | 73 | 65 | 74 | 63 | 360 | 322 | 365 | 323 |
| | 99% | 100 | 86 | 101 | 85 | 469 | 421 | 476 | 423 |
| | baserun | 44 | 38 | 44 | 38 | 164 | 146 | 167 | 148 |

We now seek to compare the **2033 results** to those from the ECA method (which is for full Phase 1). For this comparison to be meaningful, several adjustments to the results are required due to the discrepancy between the figures and methodology of the two methods. First, while our study uses HSR annual ridership (demand) forecast directly (which is 35.6 million in 2033), the ridership used by ECA method is based on planned HSR hourly capacity, which is 66.2 million passengers. Second, the ECA method assumes that 20% of HSR ridership comes from air passenger diversion; while in our study, only 5.6% (in 2029) and 7.2% (in 2023) of HSR ridership are assumed to be from air passenger diversion. Since the cost estimate of the ECA method is approximately proportional to the ridership, we scale down the ECA estimate by a factor of 66.2/35.6*20%/7.2%=5.2 to make the two estimates more comparable. Third, the ECA estimate is a lump sum of airport expansion cost across multiple years in 2018 dollars, while our method evaluates the benefit realized in future years in the unit of delay cost savings per year. Therefore, we need to calculate the future value of the ECA cost and annualize it. To do that, we first assume the expansion to occur in a specific middle year (e.g., year 2027) of Phase 1 construction. Then we estimate the future value of the ECA cost by applying an interest rate of 7%. Last, to annualize the ECA estimate, we assume a 50-year lifetime and an interest rate of 7% and obtain 504 million 2018 dollars. This is somewhat larger, but comparable to, our delay savings estimate for 2033.

## VI. CONCLUSIONS AND FUTURE WORK

This study proposes a method that estimates flight delay cost savings resulting from passengers shifting from air to HSR. We first forecast the future air capacity and demand of three major California airports, with and without HSR impact, based on a variety of data sources. Then we apply queuing diagrams to translate capacity and demand forecasts into queuing delays, which is used as a feature to predict the airport delay of Californian airports as well as all major airports in the NAS. The difference between predicted delays with and without HSR is calculated as the delay saving resulting from HSR introduction. Finally, we monetize the delay saving to compare it with the results of the ECA method. We estimate the airport delay cost savings to be 51-88 million 2018 dollars in 2029 and 235-392 million 2018 dollars in 2033. After considering the differences between our method and the ECA method, the two estimates are of a similar order of magnitude. This finding indicates that the cost of airport upgrade is approximately the same as the cost of additional delay without the upgrade, which can be used to justify the need for infrastructure investment. Besides, the rough consistency of the two delay cost saving estimates given by the NAS model and the same of SFO, LAX and SAN model supports the finding from some previous studies [29] that the impact of the congestion at large airports on the delay elsewhere in the NAS, though exists, is rather limited.

A unique feature of our analysis is that it captures day-to-day variability in flight delay and flight delay savings from HSR. We find that the flight reductions from HSR traffic diversion have a disproportionate impact on a small number of delays when airport operating conditions are very unfavorable. In our analysis, these days, while featuring high delays with or without HSR, will be substantially improved under the latter scenario. This effect is particularly pronounced at SFO. It should be noted that HSR will provide a travel option that will be particularly attractive when airport delays are high. There is a need for further research on how to best realize the potential of HSR on days when airports have low capacity and high delay.

This study has several other limitations, which can be further improved in the future. First, the CAHSR Business Plan, FAA traffic forecasts, and flight operational data all come from the pre-pandemic era, and all need to revisited in light of that calamity. Second, due to the lack of detailed capacity improvement information, we assume the same capacity for 2029 and 2033 as specified in the Airport Improvement Plan, which is likely to overestimate the delay level in 2033. Third, the Lasso model is trained using 2010-2019 data and may not be very well generalized to future scenarios if the conditions are significantly changed. In addition, the assumed linear relationship between delay and cost may be oversimplified, considering that multiple studies [30, 31] have found such relationship is not linear, unless within a quite limited delay duration domain. Last but not least, there are yet still many benefits resulting from the passenger shift from air to rail that are not captured by this study, for example, the schedule reliability, the pollution, and emission reduction, which can be further investigated in the future.

## REFERENCES

[1] Air Carrier Statistics database, T-100 Domestic Segment. US Department of Transportation, Bureau of Transportation Statistics,

[2] Official Aviation Guide. (2019). Busiest Routes 2019.

[3] Eno Center for Transportation. (2013). Addressing Future Capacity Needs in the U.S. Aviation System

[4] Hall, R. (2006). The San Diego region's air transportation future (PDF Document). Retrieved from http://sdapa.org/download/RyanHall.pdf

[5] California High-Speed Rail Authority. (2019). 2019 Equivalent Capacity Analysis Report.

[6] California High-Speed Rail Authority. (2021). 2020 Business Plan

[7] Division of Transportation Planning, California Department of Transportation. (2016). Interregional Transportation Strategic Plan.

[8] Tudela, A., Akiki, N., and Cisternas, R. (2006). Comparing the output of cost benefit and multicriteria analysis, an application to urban transport investments. Transportation Res Part A, 10, 414-423.

[9] De Rus, Ginés, and Vicente Inglada. "Cost-benefit analysis of the high-speed train in Spain." *The annals of regional science* 31.2 (1997): 175-188.

[10] Tao, Ran, et al. "Cost-benefit analysis of high-speed rail link between Hong Kong and Mainland China." *Journal of Engineering, Project, and Production Management* 1.1 (2011): 36.

[11] Levinson, David, et al. "The full cost of intercity transportation-a comparison of high speed rail, air and highway transportation in California." (1996).

[12] D'Alfonso, Tiziana, Changmin Jiang, and Valentina Bracaglia. "Would competition between air transport and high-speed rail benefit environment and social welfare?." *Transportation Research Part B: Methodological* 74 (2015): 118-137.

[13] Adler, Nicole, Eric Pels, and Chris Nash. "High-speed rail and air transport competition: Game engineering as tool for cost-benefit analysis." *Transportation Research Part B: Methodological* 44.7 (2010): 812-833.

[14] Behrens, Christiaan, and Eric Pels. "Intermodal competition in the London–Paris passenger market: High-Speed Rail and air transport." *Journal of Urban Economics* 71.3 (2012): 278-288.

[15] Yang, Hangjun, and Anming Zhang. "Effects of high-speed rail and air transport competition on prices, profits and welfare." *Transportation Research Part B: Methodological* 46.10 (2012): 1322-1333.

[16] Albalate, Daniel, Germà Bel, and Xavier Fageda. "Competition and cooperation between high-speed rail and air transportation services in Europe." *Journal of transport geography* 42 (2015): 166-174.

[17] López-Pita, Andrés, and Francesc Robusté. "The effects of high-speed rail on the reduction of air traffic congestion." *Journal of Public Transportation* 6.1 (2003): 3.

[18] Zhang, Fangni, Daniel J. Graham, and Mark Siu Chun Wong. "Quantifying the substitutability and complementarity between high-speed rail and air transport." *Transportation Research Part A: Policy and Practice* 118 (2018): 191-215.

[19] Li, Tao, and Lili Rong. "Resilience of air transport network with the complementary effects of high-speed rail network." *2019 IEEE 19th International Conference on Software Quality, Reliability and Security Companion (QRS-C)*. IEEE, 2019.

[20] Pagliara, F., Vassallo, J. M., & Román, C. (2012). High-Speed Rail Versus Air Transportation: Case Study of Madrid–Barcelona, Spain. Transportation Research Record, 2289(1), 10-17.

[21] Cao, Jing, Xiaoyue Cathy Liu, Yinhai Wang, and Qingquan Li. "Accessibility impacts of China's high-speed rail network." *Journal of Transport Geography* 28 (2013): 12-21.

[22] Dai, Lu, Mark Hansen, Michael O. Ball, and David J. Lovell. "Having a Bad Day? Predicting High Delay Days in the National Airspace System." In *Proceedings of the Fourteenth USA/Europe Air Traffic Management Research and Development Seminar (ATM2021), Virtual Event*, pp. 20-23. 2021.

[23] FAA Operations & Performance Data. Federal Aviation Administration.

[24] Cambridge Systematics. (2020) California High-Speed Rail 2020 Business Plan Ridership and Revenue Forecasting Technical Report.

[25] Cambridge Systematics. (2016) California High-Speed Rail Ridership and Revenue Model Documentation.

[26] Zhang, Yu, Monica Menendez, and Mark Hansen. "Analysis of de-peaking strategies implemented by American Airlines: causes and effects." 83rd Transportation Research Annual Meeting. 2004.

[27] Dai, Lu, Yulin Liu, and Mark Hansen. "Modeling go-around occurrence using principal component logistic regression." Transportation Research Part C: Emerging Technologies 129 (2021): 103262.

[28] FAA Office of Aviation Policy and Plans, U.S. Federal Aviation Administration. (2021) Economic Values for FAA Investment and Regulatory Decisions, a Guide: 2021 Update.

[29] Hao, Lu, Mark Hansen, Yu Zhang, and Joseph Post. "New Yoir, New York: Two ways of estimating the delay impact of New York airports." Transportation Research Part E: Logistics and Transportation Review 70 (2014):245-260.

[30] Cook, Andrew J., and Graham Tanner. "European airline delay cost reference values." (2011).

[31] Gurtner, Gérald, Andrew Cook, Anne Graham, and Samuel Cristóbal. "The economic value of additional airport departure capacity." Journal of Air Transport Management 69 (2018): 1-14.